\documentclass[aps,prb,showpacs,groupedaddress,amssymb,twocolumn]{revtex4}
\usepackage{bm,amsmath,graphicx}
\begin{document}

\title{Infrared catastrophe and tunneling into strongly correlated electron systems: Exact solution of the x-ray edge limit for the 1D electron gas and 2D Hall fluid}
\author{Kelly R. Patton and Michael R. Geller}
\affiliation{Department of Physics and Astronomy, University of Georgia,  Athens, Georgia 30602-2451}

\date{\today}

\begin{abstract}
In previous work we have proposed that the non-Fermi-liquid spectral properties in a variety of low-dimensional and strongly correlated electron systems are caused by the infrared catastrophe, and we used an exact functional integral representation for the interacting Green's function to map the tunneling problem onto the x-ray edge problem, plus corrections. The corrections are caused by the recoil of the tunneling particle, and, in systems where the method is applicable, are not expected to change the qualitative form of the tunneling density of states (DOS). Qualitatively correct results were obtained for the DOS of the 1D electron gas and 2D Hall fluid when the corrections to the x-ray edge limit were neglected and when the corresponding Nozi\`eres-De Dominicis integral equations were solved by resummation of a divergent perturbation series. Here we reexamine the x-ray edge limit for these two models by solving these integral equations exactly, finding the expected modifications of the DOS exponent in the 1D case but finding no changes in the DOS of the 2D Hall fluid with short-range interaction. We also provide, for the first time, an exact solution of the Nozi\`eres-De Dominicis equation for the 2D electron gas in the lowest Landau level.
\end{abstract}

\maketitle

\section{INTRODUCTION}\label{introduction section}

In a previous paper,\cite{Patton&GellerPRB05} we proposed a connection between anomalies in the tunneling density of states (DOS) at the Fermi energy of a wide variety of low-dimensional and strongly correlated conductors, and the infrared catastrophe. The latter is a well-known singular screening response of an ordinary metal to the sudden appearance of a localized potential, in this case produced by an electron added to the system during a tunneling event. Systems where we expect this connection to apply include all 1D electron systems,\cite{TomonagaPTP50,MattisJMP65,DzyaloshinskiiJETP74,HaldaneJPC81,HaldanePRL81,KanePRB92,Fisher97} the 2D diffusive metal\cite{AltshulerPRL80,Altshuler85,LeeRMP85,NazarovJETP89,Nazarov90,RudinPRB97,Levitov97,KhveshchenkoPRB98,KopietzPRL98,KamenevPRB99,ChamonPRB99,RollbuhlerPRL03} and Hall fluid,\cite{EisensteinPRL92,YangPRL93,HatsugaiPRL71,HePRL93,JohanssonPRL93,KimPRB94,AleinerPRL95,HaussmannPRB96,LeonardJPCM98,WangPRL99} and the edge of the confined Hall fluid.\cite{WenPRL90,WenPRB90a,WenPRB91a,MoonPRL93,KanePRL94,KanePRB95,MillikinSSC96,ChangPRL96,HanPRB97a,HanPRB97b,Conti97,GraysonPRL98,ShytovPRL98,ContiJPCM98,LopezPRB99,ZulickePRB99,PruiskenPRB99,KhveshchenkoSSC99,MoorePRB00,AlekseevJETP00,ChangPRL01,GoldmanPRL01,HilkePRL01,LopezPRB01,PasquierPRB01,TsiperPRB01,LevitovPRB01,MandalSSC01,WanPRL02,MandalPRL02,ZulickePRB02,ChangRMP03,HuberPRL05} We argued that in such systems, the accommodation of a new electron added during a tunneling event is frustrated by the low dimensionality, or the localizing effects of a magnetic field or disorder, or both. In these cases the tunneling problem is similar to the x-ray edge problem. 

A mapping between the two is made via an exact scalar functional integral representation for the interacting propagator, which replaces it by a Gaussian average of noninteracting propagators for electrons in the presence of potentials $\phi({\bf r},\tau)$. We then singled out a dangerous field configuration
\begin{equation}
\phi_{\rm xr}({\bf r},\tau)=U({\bf r})\Theta(\tau_{0}-\tau)\Theta(\tau),
\label{phi xray}
\end{equation}
which would be the potential produced by an electron added to the origin at time $\tau = 0$ and removed at a later time $\tau_0,$ if it had an infinite mass.  Here $U({\bf r})$ is bare electron-electron interaction. In Ref.~[\onlinecite{Patton&GellerPRB05}] this special field configuration was treated by resumming a divergent series, and fluctuations about $\phi_{\rm xr}({\bf r},\tau)$ were ignored, yet qualitatively correct expressions for the DOS were obtained.

To obtain quantitatively correct results it will be necessary to go beyond this ``perturbative" x-ray edge limit. In Ref.~[\onlinecite{Patton&GellerPre05}] we proposed and investigated a functional cumulant expansion method that includes field fluctuations away from $\phi_{\rm xr}({\bf r},\tau)$, and treats field configurations close to $\phi_{\rm xr}({\bf r},\tau)$ perturbatively as in Ref.~[\onlinecite{Patton&GellerPRB05}]. Although the improved method yields the exact DOS exponent for the important Tomonaga-Luttinger model, calculable by bosonization, we do not expect it to be generally exact in 1D. (Furthermore, the method fails in the presence of a strong magnetic field because of the ground state degeneracy.) In this paper we neglect fluctuations about $\phi_{\rm xr}({\bf r},\tau)$ but treat that field configuration exactly (in the relevant long $\tau_0$ asymptotic limit). This is accomplished by finding the exact low-energy solution of the Dyson equation for noninteracting electrons in the presence of $\phi_{\rm xr}({\bf r},\tau)$, which we refer to as the Nozi\`eres-De Dominicis equation. We carry out this analysis for the 1D electron and 2D Hall fluid, both with short range interaction. To this end we obtain, for the first time, an exact solution of the Nozi\`eres-De Dominicis equation for the 2D electron gas in the lowest Landau level.

\section{Formalism}

We calculate the tunneling DOS by analytic continuation of the Euclidean propagator
\begin{equation}
G({\bf r}_{\rm f}\sigma_{\rm f},{\bf r}_{\rm i} \sigma_{\rm i},\tau_0) \equiv - \big\langle T \psi_{\sigma_{\rm f}}({\bf r}_{\rm f} , \tau_0) {\bar \psi}_{\sigma_{\rm i}}({\bf r}_{\rm i},0)
\big\rangle_{\! \scriptscriptstyle H},
\label{G definition}
\end{equation}
where $H = H_0 + V$ is the grand-canonical Hamiltonian for the $D$-dimensional interacting electron system, with
\begin{equation*}
H_0 \equiv \sum_\sigma \! \int \! d^{\scriptscriptstyle D}r \, \psi^\dagger_\sigma({\bf r}) \, \bigg[{\Pi^2 \over 2m} + v({\bf r}) - \mu \bigg] \psi_\sigma ({\bf r})
\label{H0}
\end{equation*}
and
\begin{equation*}
V \equiv {\textstyle{1 \over 2}} \int \! d^{\scriptscriptstyle D}r \, d^{\scriptscriptstyle D} r' \, \delta n({\bf r}) \, U({\bf r}-{\bf r}') \, \delta n({\bf r}').
\label{V}
\end{equation*}
Here ${\bf \Pi} \equiv {\bf p} + {\textstyle{e \over c}} {\bf A},$ with ${\bf A}$ the vector potential (if a magnetic field is present), and
\begin{equation}
\delta n({\bf r}) \equiv \sum_\sigma \psi_\sigma^\dagger({\bf r})  \psi_\sigma({\bf r}) - n_0({\bf r})
\end{equation}
is the density fluctuation operator. $H_0$ is the Hamiltonian in the Hartree approximation, and $v({\bf r})$ includes  any single-particle potential along with the Hartree interaction with the self-consistent density $n_0({\bf r})$.

After an exact Hubbard-Stratonovich transformation, the interacting Green's function can be written as functional integral over a scalar field $\phi,$
\begin{eqnarray}
&&G({\bf r}_{\rm f} \sigma_{\rm f}, {\bf r}_{\rm i} \sigma_{\rm i},\tau_0) = {\cal N} e^{{1 \over 2} \! \int \! \phi_{\rm xr} U^{-1} \phi_{\rm xr}} \nonumber \\
& \times & \int \! D \mu[{\phi}] \, e^{-i \! \int \! \phi U^{-1} \phi_{\rm xr}} \,  g({\bf r}_{\rm f} \sigma_{\rm f}, {\bf r}_{\rm i} \sigma_{\rm i},\tau_0 | i\phi_{\rm xr} + \phi),  {\hskip 0.1in} 
\label{shifted functional integral}
\end{eqnarray} 
where ${\cal N} \! \equiv \! \langle T \exp(-\int_0^\beta d \tau \, V) \rangle_0^{-1}$ is a constant, independent of  $\tau_0$, 
\begin{equation}
D\mu[{\phi}] \equiv \frac{D\phi \, e^{-{1 \over 2} \int \phi U^{-1} \phi}}{ \int \! D\phi \ e^{-{1 \over 2} \int \phi U^{-1} \phi}}
\end{equation}
is a measure normalized according to $\int \! D\mu[{\phi}] = 1,$ and
\begin{eqnarray}
&&g({\bf r}_{\rm f} \sigma_{\rm f}, {\bf r}_{\rm i} \sigma_{\rm i},\tau_0 | \phi) \nonumber \\
&\equiv& \! \! - \big\langle T \psi_{\sigma_{\rm f}}({\bf r}_{\rm f},\tau_0)  {\bar \psi}_{ \sigma_{\rm i}}({\bf r}_{\rm i},0) \, e^{i \int_0^\beta \! d \tau \int \! d^{\scriptscriptstyle D}r \, \phi({\bf r},\tau) \, \delta n({\bf r},\tau) } \big\rangle_0 {\hskip 0.20in}
\label{g correlation function definition}
\end{eqnarray}
is a noninteracting functional of $\phi$. 

The dangerous field configuration $\phi_{\rm xr}({\bf r},\tau)$, which itself depends on the parameters ${\bf r}_{\rm i},$ $ {\bf r}_{\rm f},$ and $\tau_0$ appearing in (\ref{G definition}), has been given in Ref.~[\onlinecite{Patton&GellerPRB05}]. For the case the tunneling DOS at point ${\bf r}_0$ of interest here we have definition (\ref{phi xray}), which is the potential that would be produced by the added particle in (\ref{G definition}) if it had an infinite mass. Fluctuations about $\phi_{\rm xr}$ account for the recoil of the finite-mass tunneling electron. 

In the x-ray edge limit, we ignore fluctuations about $\phi_{\rm xr},$ in which case 
\begin{equation}
G({\bf r}_{\rm f} \sigma_{\rm f}, {\bf r}_{\rm i} \sigma_{\rm i},\tau_0) \approx {\cal N} g({\bf r}_{\rm f} \sigma_{\rm f}, {\bf r}_{\rm i} \sigma_{\rm i},\tau_0| i\phi_{\rm xr}).
\label{propagator in xray limit}
\end{equation}
Eq.~(\ref{propagator in xray limit}) defines the interacting propagator in the x-ray edge limit. The local tunneling DOS at position ${\bf r}_0$ is obtained by setting ${\bf r}_{\rm i} = {\bf r}_{\rm f} = {\bf r}_0$ and $\sigma_{\rm i} = \sigma_{\rm f} = \sigma_0$, and summing over $\sigma_0$. In the remainder of this paper we will evaluate (\ref{propagator in xray limit}) for the 1D electron gas and the 2D Hall fluid, with a short-range interaction of the form
\begin{equation} 
U({\bf r}) = \lambda \delta({\bf r}).
\label{short range interaction definition}
\end{equation}

\section{X-RAY GREEN'S FUNCTION}\label{dyson section}

The quantity $g({\bf r}_{\rm f} \sigma_{\rm f}, {\bf r}_{\rm i} \sigma_{\rm i},\tau_0| i\phi_{\rm xr})$ required in (\ref{propagator in xray limit}) is related to the Euclidean propagator,
\begin{equation*}
G_{\rm xr}({\bf r} \sigma \tau , {\bf r}' \sigma' \tau') \equiv  - \frac{ \langle T \psi_{ \sigma}({\bf r} ,\tau) {\bar \psi_{\sigma'}}({\bf r}' ,\tau') e^{- \! \int \phi_{\rm xr} \delta n} \rangle_0} {\langle T e^{- \! \int \phi_{\rm xr} \delta n}  \rangle_0},
\label{Gxr definition}
\end{equation*}
according to
\begin{equation}
g({\bf r}_{\rm f} \sigma_{\rm f}, {\bf r}_{\rm i} \sigma_{\rm i},\tau_0| i\phi_{\rm xr})  = G_{\rm xr}({\bf r}_{\rm f} \sigma_{\rm f} \tau_0, {\bf r}_{\rm i} \sigma_{\rm i} 0) \, Z_{\rm xr}(\tau_0) ,
\label{g in terms of Gxr}
\end{equation}
with
\begin{equation}
Z_{\rm xr}(\tau_0) \equiv \langle T e^{- \int_0^\beta d \tau \int d^{\scriptscriptstyle D}r \, \phi_{\rm xr}({\bf r},\tau) \, \delta n({\bf r},\tau) } \rangle_0 .
\label{Zxr definition}
\end{equation}
We refer to $G_{\rm xr}({\bf r} \sigma \tau , {\bf r}' \sigma' \tau')$ as the x-ray Green's function which describes noninteracting electrons in the presence of a real-valued potential $\phi_{\rm xr}({\bf r},\tau)$. It satisfies the Dyson equation  
\begin{align}
&G_{\rm xr}({\bf r} \sigma \tau , {\bf r}' \sigma \tau') = G_0({\bf r} \sigma , {\bf r}' \sigma, \tau- \tau') \nonumber  \\
&+ \int d^{\scriptscriptstyle D}{\bar r} \, d{\bar \tau} \  G_0({\bf r} \sigma , {\bar {\bf r}} \sigma, \tau- {\bar \tau}) \, \phi_{\rm xr}({\bar {\bf r}},{\bar \tau}) \, G_{\rm xr}({\bar {\bf r}} \sigma {\bar \tau} , {\bf r}' \sigma \tau'). {\hskip 0.3in}
\label{general Gxr dyson equation}
\end{align}
Here we have used that fact that the x-ray Green's function is diagonal in spin. For a calculation of the DOS we use the form (\ref{phi xray}), in which case (\ref{general Gxr dyson equation}) becomes
\begin{eqnarray}
&&G_{\rm xr}({\bf r} \sigma \tau , {\bf r}' \sigma \tau') = G_0({\bf r} \sigma , {\bf r}' \sigma, \tau- \tau') \nonumber  \\
&+& \lambda \int_0^{\tau_0} \! dt \ G_0({\bf r} \sigma , {\bf r}_0 \sigma, \tau- t) \, G_{\rm xr}({\bf r}_0 \sigma t , {\bf r}' \sigma \tau'), {\hskip 0.3in}
\label{Gxr dyson equation}
\end{eqnarray}
where we have assumed the short-range interaction (\ref{short range interaction definition}).

By using the linked cluster expansion and coupling-constant integration, $Z_{\rm xr}$ can be shown to be related to the x-ray Green's function by\cite{Nozieres&DeDominicisPR69}
\begin{equation}
Z_{\rm xr}(\tau_0) = e^{n_0 \lambda \tau_0} e^{ \! -\lambda \sum_\sigma \int_{0}^{1} \! \! d \xi \int_0^{\tau_0} \! \! d\tau \ G^{\xi}_{\rm xr}({\bf r}_0 \sigma \tau , {\bf r}_0 \sigma \tau^+)},
\label{Zxr identity}
\end{equation}
where $G^{\xi}_{\rm xr}({\bf r} \sigma \tau , {\bf r}' \sigma \tau')$ is the solution of (\ref{Gxr dyson equation}) with scaled coupling constant $\xi \lambda$. 

There is no ${\bf r}_0$ dependence in the DOS for the translationally invariant models considered here and we can take ${\bf r}_0 = 0.$

\section{1D ELECTRON GAS}\label{electron gas section}

 $G_{\rm xr}(0 \sigma \tau , 0 \sigma \tau')$  was calculated exactly in the large $\tau_0$, asymptotic limit for the 3D electron gas in zero field by Nozi\`{e}res and De Dominicis.\cite{Nozieres&DeDominicisPR69} Their result is actually valid for arbitrary spatial dimension $D$ if the appropriate noninteracting DOS is used. 

We take the asymptotic form of the noninteracting propagator as
\begin{equation}
G_0(\tau) \approx -  \, {\rm P} \, \frac{N_0}{\tau}, \ \ \  {\rm with} \ \ \  N_0 \equiv \frac{1}{\pi v_{\rm F}}.
\label{1D electron gas P regularized G0}
\end{equation}
$N_0$ is the noninteracting DOS per spin component at $\epsilon_{\rm F},$ and ${\rm P}$ denotes the principal part. The solution of (\ref{Gxr dyson equation}) for this model with ${\bf r}={\bf r}'=0$ is
\begin{align}
G_{\rm xr}(\tau_0)=-&N_0\cos(\delta_{\lambda})\left[P\frac{\cos(\delta_{\lambda})}{\tau_0}+\pi\sin(\delta_{\lambda})\delta(\tau_0)\right]\nonumber \\ &\times
\left(\frac{a}{\tau_0}\right)^{2\delta_{\lambda}/{\pi}},
\end{align}
and
\begin{equation}
Z_{\rm xr}(\tau_0)=\left(\frac{a}{\tau_0}\right)^{2\left(\delta_{\lambda}/{\pi}\right)^2},
\end{equation}
where $\delta_\lambda$ is the scattering phase shift of the electrons caused by the potential  $\phi_{\rm xr}$ given by
\begin{equation}
\delta_{\lambda}={\rm arctan} (N_0\pi \lambda)
\end{equation}
and $a$ is a short time cut-off on the order of the Fermi energy. 

Thus 
\begin{equation}
G(\tau_0)\approx g(\tau_0|i\phi_{\rm xr})\sim\left(\frac{1}{\tau_0}\right)^{1+2\delta_{\lambda}/{\pi}+2\left(\delta_{\lambda}/{\pi}\right)^2}
\end{equation}
which gives a DOS in the x-ray edge limit as
\begin{equation}
N(\epsilon)\sim \epsilon^{2\delta_{\lambda}/{\pi}+2\left(\delta_{\lambda}/{\pi}\right)^2}.
\label{1d dos}
\end{equation}
By expanding the exponent in (\ref{1d dos}) in powers of the coupling paramenter $\lambda$ one recovers the perturbative x-ray result of Ref.~[\onlinecite{Patton&GellerPRB05}].

\section{2D HALL FLUID}\label{hall fluid section}

\begin{figure}
\includegraphics[width=8.0cm]{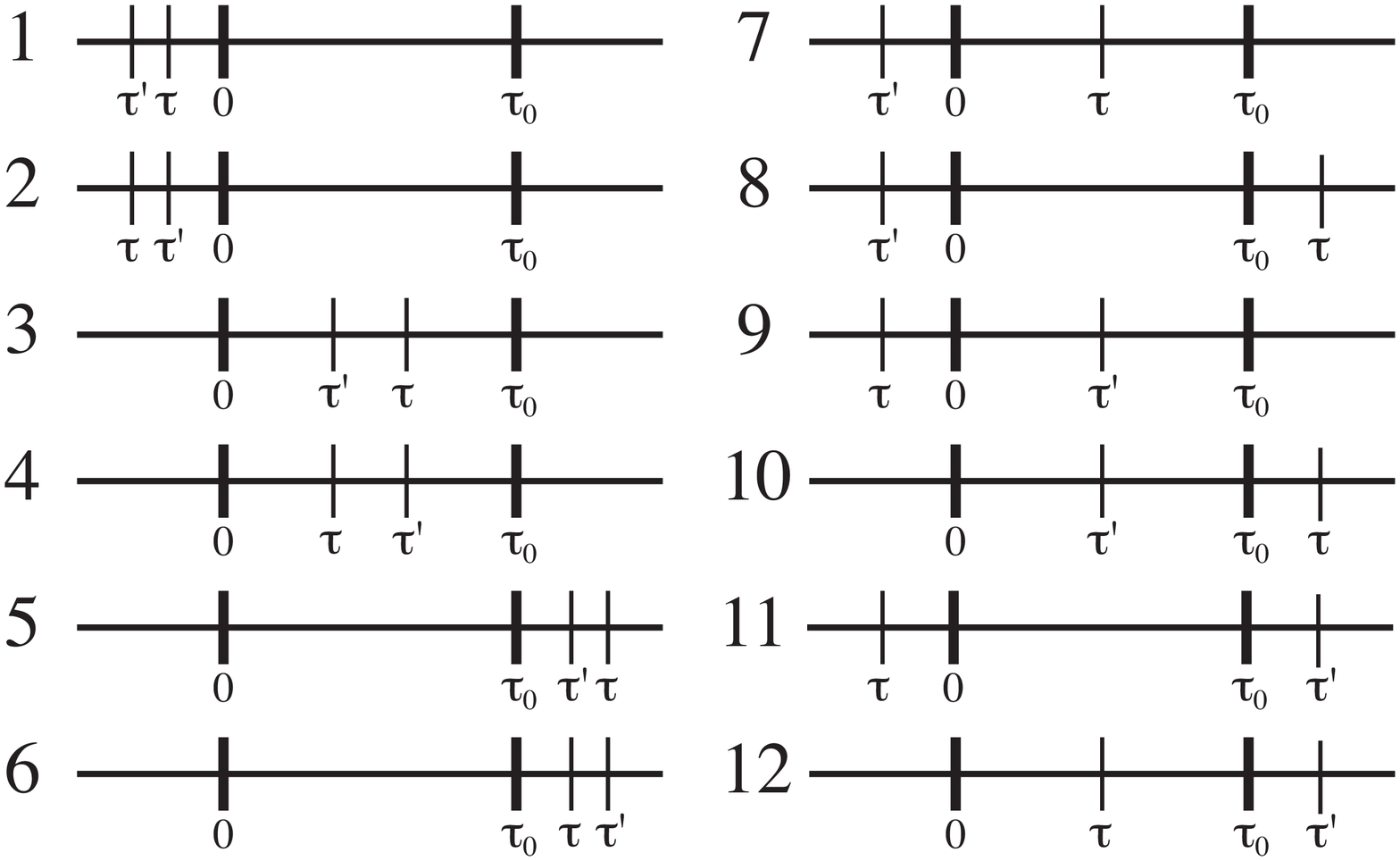}
\caption{The 12 possible time orderings $k=1,2,\cdots \! ,12$ of $G_{\rm xr}(0 \tau,0 \tau').$ $\tau_0$ is assumed to be nonnegative.}
\label{time ordering figure}
\end{figure} 

Unlike the low-energy Dyson equation  for the 1D electron gas, which is solvable by Hilbert transform techniques, there are no standard methods available to solve the corresponding integral equation for the Hall fluid. We were able to guess the exact analytic solution, aided by perturbation theory and by numerical studies carried out by expansion in a plane-wave basis followed by matrix inversion. 

We assume the system to be spin-polarized and spin labels are suppressed. In the Landau gauge ${\bf A} = B x {\bf e}_y$, the noninteracting propagator in the $|\tau| \gg \omega_{\rm c}^{-1}$ limit is
\begin{equation}
G_0({\bf r},{\bf r}',\tau) =  \, \Gamma({\bf r},{\bf r}') \, [ \nu -  \Theta(\tau)],
\label{LLL G0}
\end{equation}
where $\nu$ is the filling factor satisfying  $0 \le \nu \le 1$, and
\begin{equation}
\Gamma({\bf r},{\bf r}') \equiv  {1 \over 2 \pi \ell^2} \, e^{-|{\bf r}-{\bf r}'|^2 / 4 \ell^2} \, e^{-i(x+x')(y-y')/2\ell^2}.
\end{equation}

First consider the case where ${\bf r}_{\rm i} = {\bf r}_{\rm f} = {\bf r}_0.$ We can let $ {\bf r}_0 = 0$ without loss of generality and at the origin (\ref{Gxr dyson equation}) reduces to
\begin{eqnarray}
&&G_{\rm xr}(0 \tau,0 \tau') = \frac{\nu - \Theta(\tau - \tau')}{2 \pi \ell^2}  \nonumber \\
&+& \gamma \int_{0}^{\tau_0} \! \! dt \, \big[ \nu - \Theta(\tau-t)\big]  \, G_{\rm xr}(0t,0\tau'),
\label{Gxr origin equation}
\end{eqnarray}
where
\begin{equation}
\gamma \equiv { \lambda \over 2 \pi \ell^2}
\end{equation}
is an interaction strength with dimensions of energy.

The time arguments of $G_{\rm xr}(0 \tau,0 \tau')$ on the left side of Eq.~(\ref{Gxr origin equation}) can assume the 12 possible orderings $k=1,2,\cdots,12$ defined in Fig.~\ref{time ordering figure}; the right side produces terms with two or more different orderings $k', k'', \cdots$. We therefore seek a solution of the form
\begin{equation}
G_{\rm xr}(0 \tau,0 \tau') = \sum_{k} A_k \, W_k(\tau,\tau') \, f_k({\tau}),
\label{trail solution}
\end{equation} 
where $W_k(\tau,\tau')$ is unity if $\tau$ and $\tau'$ have ordering $k$ and zero otherwise; an explicit form for $W_k(\tau,\tau')$ is given in Appendix \ref{W appendix}. The functions $f_k({\tau})$ are chosen to reflect the fact that an electron accumulates an additional phase $\gamma \, \Delta \tau$ while in the presence of $\phi_{\rm xr}$ for a time $\Delta \tau$, whereas a hole acquires a phase $- \gamma \, \Delta \tau$. The 12 unknown coefficients $A_k$ (which depend parametrically on $\tau_0$ and $\tau'$) are obtained by solving the 12 linearly independent equations resulting from the decomposition of (\ref{Gxr origin equation}) into distinct time orderings $k=1,2,\cdots,12.$ The result is
\begin{widetext}
\begin{align}
G_{\rm xr}(0 \tau,0 \tau') &= \frac{1}{2 \pi \ell^2} \bigg({1 \over 1 - \nu + \nu e^{- \gamma \tau_0}}\bigg) 
\bigg[ (\nu-1) \, \big( W_1 + W_5 \big) + \nu  e^{- \gamma \tau_0} \, \big( W_2 + W_6 \big) + (\nu-1) e^{- \gamma (\tau-\tau')} \, W_3 
\nonumber \\ &+ \nu e^{- \gamma \tau_0} e^{ \gamma (\tau'-\tau)} \, W_4 + (\nu-1)  e^{- \gamma \tau} \, W_7 + (\nu-1)  e^{- \gamma \tau_0} \, W_8 
+ \nu  e^{-\gamma(\tau_0-\tau')} \, W_9 + (\nu-1)  e^{- \gamma (\tau_0-\tau')} \, W_{10} \nonumber \\ &+ \nu \, W_{11} + \nu  e^{-\gamma \tau} \, W_{12} \bigg].
\end{align}
As a side note, the solution for general ${\bf r}_{\rm i}$ and ${\bf r}_{\rm f}$ is obtained using the same method and when both $\tau$ and $\tau'$ are in the interval $(0,\tau_0),$ the result is
\begin{eqnarray}
G_{\rm xr}({\bf r}\tau,{\bf r}' \tau') &=& [\nu \! - \! \Theta(\tau-\tau')]\big[ \Gamma({\bf r},{\bf r}') - 2 \pi \ell^2 \, \Gamma({\bf r},0) \, \Gamma(0,{\bf r}') \big] \nonumber \\
&+& 2 \pi  \ell^2 \,  \Gamma({\bf r},0) \, \Gamma(0,{\bf r}') \bigg[  {(\nu-1) \, \Theta(\tau-\tau') + \nu e^{- \gamma \tau_0} \, \Theta(\tau'-\tau) \over 1 - \nu + \nu e^{- \gamma \tau_0} }\bigg] e^{-\gamma(\tau-\tau')};
\label{Gxr}
\end{eqnarray}
the other cases follow similarly. At the origin (\ref{Gxr}) reduces to
\begin{equation}
G_{\rm xr}(0 \tau,0 \tau') = {1 \over 2 \pi \ell^2}  \bigg[  {(\nu-1) \, \Theta(\tau-\tau') + \nu e^{- \gamma \tau_0} \, \Theta(\tau'-\tau) \over 1 - \nu + \nu e^{- \gamma \tau_0} }\bigg] e^{-\gamma(\tau-\tau')}.
\label{Gxr at origin}
\end{equation}
\end{widetext}
Finally
\begin{equation}
G_{\rm xr}(0 \tau_0,00) = {1 \over 2 \pi \ell^2} \bigg(\frac{\nu-1}{1 - \nu + \nu e^{- \gamma \tau_0}} \bigg) e^{-\gamma \tau_0}.
\label{Gxr for g}
\end{equation}
Using Eq.~(\ref{Zxr identity}) we obtain 
\begin{equation}
Z_{\rm xr} = e^{\nu \gamma \tau_0}(1 - \nu + \nu e^{- \gamma \tau_0}),
\end{equation}
therefore 
\begin{equation}
g({\bf r}0 \sigma_0, {\bf r}_0 \sigma_0,\tau_0| i\phi_{\rm xr}) = \frac{\nu -1}{2 \pi \ell^2} \, e^{\gamma (\nu-1) \tau_0}.
\end{equation}
This is identical to what we obtained in Ref.~[\onlinecite{Patton&GellerPRB05}]. The tunneling DOS is therefore
\begin{equation}
N(\epsilon) = {\rm const} \! \times \! \delta\big(\epsilon - [1-\nu]\gamma \big).
\label{hall fluid xray DOS}
\end{equation}

\newpage

\section{discussion}\label{discussion section}

In this paper, we have carried out an exact treatment of the x-ray edge limit introduced in Ref.~[\onlinecite{Patton&GellerPRB05}], for the same models considered there. Whereas the 1D electron gas result (\ref{1d dos}) would be expected, the DOS of the 2D Hall fluid remains gapped as in Ref.~[\onlinecite{Patton&GellerPRB05}]. A generalization of our method that accounts for fluctuations about $\phi_{\rm xr}$, and that can be used in a magnetic field, will be needed to recover the actual pseudogap of the Hall fluid.\cite{EisensteinPRL92,YangPRL93,HatsugaiPRL71,HePRL93,JohanssonPRL93,KimPRB94,AleinerPRL95,HaussmannPRB96,LeonardJPCM98,WangPRL99}

\appendix

\section{TIME ORDERING FUNCTIONS}\label{W appendix}

Let
\begin{eqnarray}
W_1(\tau,\tau') &\equiv& \Theta(-\tau) \, \Theta(-\tau') \, \Theta(\tau - \tau') \nonumber \\
W_2(\tau,\tau') &\equiv& \Theta(-\tau) \, \Theta(-\tau') \, \Theta(\tau' - \tau) \nonumber \\
W_3(\tau,\tau') &\equiv& W(\tau) \, W(\tau') \, \Theta(\tau-\tau') \nonumber \\
W_4(\tau,\tau') &\equiv& W(\tau) \, W(\tau') \, \Theta(\tau'-\tau) \nonumber \\
W_5(\tau,\tau') &\equiv& \Theta(\tau - \tau_0) \, \Theta(\tau' - \tau_0) \, \Theta(\tau - \tau') \nonumber \\
W_6(\tau,\tau') &\equiv& \Theta(\tau - \tau_0) \, \Theta(\tau' -\tau_0) \, \Theta(\tau' - \tau) \nonumber \\
W_7(\tau,\tau') &\equiv& W(\tau) \, \Theta(-\tau') \nonumber \\
W_8(\tau,\tau') &\equiv& \Theta(\tau - \tau_0) \, \Theta(-\tau')  \nonumber \\
W_9(\tau,\tau') &\equiv&  \Theta(-\tau) \, W(\tau')  \nonumber \\
W_{10}(\tau,\tau') &\equiv& \Theta(\tau - \tau_0) \, W(\tau')  \nonumber \\
W_{11}(\tau,\tau') &\equiv& \Theta(-\tau) \, \Theta(\tau' - \tau_0) \nonumber \\
W_{12}(\tau,\tau') &\equiv& W(\tau) \, \Theta(\tau' - \tau_0), \nonumber
\end{eqnarray}
where $\Theta(t)$ is the Heaviside step function and $W$ (with no subscripts) is the a window function, defined as
\begin{equation}
W \equiv \Theta(\tau_0-\tau)\Theta(\tau).
\end{equation}

\acknowledgments

This work was supported by the National Science Foundation under Grant No.~DMR-0093217, and by a Cottrell Scholars Award from the Research Corporation. It is a pleasure to thank Phil Anderson, Matthew Grayson, Dmitri Khveshchenko,  Allan MacDonald, Emily Pritchett, David Thouless, Shan-Ho Tsai, and Giovanni Vignale for useful discussions. M.G. would also like to acknowledge the Aspen Center for Physics, where some of this work was carried out.

\bibliography{/Users/mgeller/Papers/bibliographies/MRGhall,/Users/mgeller/Papers/bibliographies/MRGmanybody,/Users/mgeller/Papers/bibliographies/MRGpre,/Users/mgeller/Papers/bibliographies/MRGbooks,/Users/mgeller/Papers/bibliographies/MRGgroup,exactnotes}

\end{document}